\documentclass[aps,pra,epsfigure,twocolumn,showpacs]{revtex4} 
\usepackage{amsmath}
\usepackage{amstext}
\usepackage{latexsym}
\usepackage{graphicx}
\usepackage{amsfonts}
\usepackage{epsfig}
\usepackage{color}

\newcommand{\Tr}{\mathrm{Tr}}

\newcommand{\petr}[1]{\textcolor{black}{#1}} 

\begin{document}
\title{Resources for universal quantum state manipulation and engineering}

\author{Petr Marek}
\affiliation{Department of Optics, Palack\'{y} University, 17. listopadu 50,
77200 Olomouc, Czech Republic}

\author{Jarom\'{\i}r Fiur\'{a}\v{s}ek}
\affiliation{Department of Optics, Palack\'{y} University, 17. listopadu 50,
77200 Olomouc, Czech Republic}

\begin{abstract}
We investigate which non-Gaussian resources are needed, in addition to Gaussian operations and measurements, for implementation of arbitrary quantum gates on multimode quantum states of light. We  show that an arbitrary set of states with finite expansion in Fock basis is sufficient for this task. As an illustration we present an explicit scheme for probabilistic implementation of the nonlinear sign gate using resource non-Gaussian states and Gaussian operations.

\end{abstract}


\maketitle

\section{Introduction}

The ability to perform an arbitrary operation on a quantum system is a crucial prerequisite for advanced quantum information processing and quantum computing \petr{\cite{Braun05}}.  In optical implementations, quantum states of light are manipulated mainly with passive and active linear optical elements such as beam splitters and squeezers. The resulting state transformations preserve the Gaussian form of the Wigner function and are thus referred to as Gaussian operations. It is readily apparent that such operations  alone are not sufficient for universal continuous-variable (CV) quantum computation \cite{Lloyd99,Bartlett02a,Bartlett02b} and must be supplemented by access to some other resources such as nonlinear dynamics \cite{Lloyd99}, single-photon detectors \cite{Knill01,Gu09}, or non-Gaussian states \cite{Gottesman01,Bartlett03}. While several schemes for generation of highly nonclassical states of light and implementation of various non-Gaussian operations have been suggested \cite{CKW,Fiurasek03,Fiurasek05}, a systematic study of usefulness of non-Gaussian states for universal quantum state manipulation and engineering has been missing.

\petr{In the present paper we focus on implementation of quantum gates using off-line generated ancilla states $|\psi\rangle$ and Gaussian measurements and operations \cite{Bartlett03,Filip05,Yoshikawa08}. The ancilla states  represent the only non-Gaussian ingredient and can thus be seen as a resource that is converted into a non-Gaussian CV quantum gate. It is our aim to investigate what non-Gaussian ancilla states are sufficient for realization of arbitrary CV quantum gate within this approach. We shall prove that arbitrary pure single-mode non-Gaussian state $|\psi\rangle$ possessing finite expansion in Fock-state basis is sufficient for (probabilistic) implementation of any $n$-mode quantum gate on Hilbert space $\mathcal{H}_N^{\otimes n }$, where $\mathcal{H}_N$ is spanned by the first $N+1$ Fock states and both $N$ and $n$ are finite but otherwise arbitrary. The formulation in terms of truncated finite-dimensional Hilbert spaces is necessary in order to ensure that a scheme with finite number of components can be constructed that (conditionally) implements the requested gate. }

The core of our argument is the reduction of the problem to generation of single-photon Fock states $|1\rangle$ from the resource state $|\psi\rangle$. We provide explicit scheme for this latter task and assess its performance.   For the sake of presentation clarity we explain the protocol on the example of traveling light modes, but the scheme is applicable also to other physical platforms such as atomic ensembles
or optomechanical systems.

\begin{figure}[!b!]
\centerline{\psfig{figure=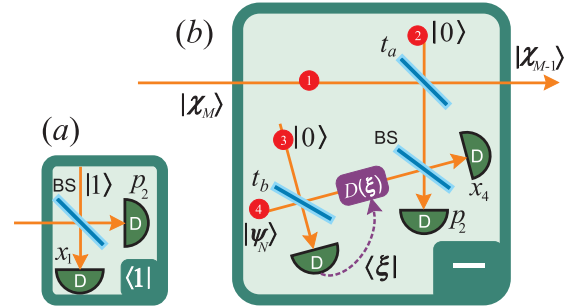,width=0.85\linewidth}}
\caption{(color online) (a) Setup for projective measurement on a single-photon state. D - Homodyne detectors, BS - balanced beam splitter.
(b) Setup for approximate photon subtraction. D - homodyne detector, BS - balanced beam splitter, $t_{a,b}$ - beam splitter with with transmittance $t_{a,b}$, $D(\xi)$ - displacement driven by detected value $\xi$.} \label{fig_singlephotondet}
\end{figure}

\section{Sufficiency of single-photon states}

We start by demonstrating that only single-photon states, apart from Gaussian operations and measurements, are required for probabilistic implementation of arbitrary quantum operation on $\mathcal{H}_N^{\otimes n}$.
A crucial observation is that the projection on a single-photon state can be performed with help of an ancillary single-photon state, a balanced beam splitter and a pair of homodyne detectors
measuring amplitude quadrature $x_1$ and phase quadrature $p_2$, respectively, c.f. Fig.~\ref{fig_singlephotondet}(a).  Successful projection is heralded by outcomes $x_1=0$ and $p_2=0$. In this case, the two input modes impinging on the balanced beam splitter are  projected on the maximally entangled
EPR state $|\Psi_{\mathrm{EPR}}\rangle=\sum_{n=0}^{\infty} |n,n\rangle$. This in conjunction with the ancillary single-photon state implements the probabilistic projection on a single-photon state,
$_{12}\langle \Psi_{\mathrm{EPR}}| 1\rangle_1=\, _2\langle 1|.$
To achieve a nonzero success probability, 
a finite acceptance window for the measurement outcomes $x_1$ and $p_2$ has to be introduced, which reduces the fidelity of the projection and leads to trade-off between operation quality and its success probability. This is an unavoidable feature of our protocol arising from involvement of only Gaussian measurements.

Single-photon states and single-photon measurements combined with Gaussian operations are sufficient for probabilistic  preparation of arbitrary multimode quantum state \cite{Fiurasek03} and implementation of arbitrary transformation on $\mathcal{H}_N^{\otimes n}$, e.g. by exploiting the scheme described in Ref. \cite{CKW} or simply by quantum teleportation \cite{Gottesman99}.
The whole question about nature of non-Gaussian resources sufficient for universal quantum state manipulation is thereby reduced to finding a class of states from which a single-photon state can be generated with help of only Gaussian operations and measurements. We are going to show that any collection of non-Gaussian pure states possessing finite expansion in the Fock-state basis is sufficient for this.

\section{Generalized photon subtraction}

Let us consider a steady supply of states of the form
\begin{equation}
   |\psi_N\rangle =  \sum_{k=0}^{N} c_k|k\rangle.
   \label{psiN}
\end{equation}
An essential ingredient of our protocol is  the setup depicted in Fig.~\ref{fig_singlephotondet}(b)
which employs one auxiliary state $|\psi_N\rangle$ and Gaussian operations
to remove the highest Fock state $|M\rangle$ from the input state $|\chi_M\rangle=\sum_{m=0}^M b_m |m\rangle$. This produces a state
$|\chi_{M-1}\rangle=\sum_{m=0}^{M-1} b_m' |m\rangle$ and this operation can be thus seen as a version of approximative photon subtraction.

First part of the process lies
in a deterministic transformation of $|\psi_N\rangle$  into a state
\begin{equation}\label{1step02}
   |\phi_{\overline{0}}\rangle = \sum_{k=1}^{\infty} d_k |k\rangle,\quad \sum_{k=1}^{\infty}|d_k|^2 = 1,
\end{equation}
with $d_1 \neq 0$ and missing vacuum term, $d_0=0$. This can be achieved by coherent displacement of the state $|\psi_N\rangle$ if the displacement amplitude $\alpha$ satisfies
\begin{equation}
    \langle 0|D(\alpha)|\psi_N\rangle =e^{-|\alpha|^2/2} \sum_{k=0}^{N} c_k \frac{(-\alpha^\ast)^k}{\sqrt{k!}} = 0.
    \label{alpharoot}
\end{equation}
Such $\alpha$ exists for all finite $N$. 
However, for a particular set of scenarios, \emph{e.g.} when $|\psi_N\rangle = |N\rangle$, this approach does not work as the required displacement is $\alpha = 0$ corresponding to no action
at all and the scheme in Fig.~\ref{fig_singlephotondet}(b) would produce vacuum state from input $|\psi_N\rangle$. This problem can be fortunately circumvented using an ancillary vacuum mode, a beam splitter, a single homodyne detection and feed-forward, see Fig.~\ref{fig_singlephotondet}(b). After passing through the beam splitter with transmittance $t_b$, the homodyne detection of the amplitude quadrature $\hat{x}_3$  yielding a value $x$, and the displacement $\alpha$, the state $|N\rangle$ transforms into
\begin{equation}
| \phi\rangle=   D(\alpha) \sum_{k=0}^N \sqrt{N\choose k}
    t_b^k r_b^{N-k}\langle x |N-k\rangle |k\rangle,
\end{equation}
where $r_j = \sqrt{1-t_j^2}$ for any $j$. By employing the relation for an overlap of a quadrature eigenstate and a Fock state,
\begin{equation}\label{fockXoverlap}
    \langle x|n\rangle = \frac{H_n(x)}{\pi^{1/4}\sqrt{n!2^n}} e^{-x^2/2},
\end{equation}
where $H_n(x)$ stands for the Hermite polynomial, we can see that to arrive at the form (\ref{1step02}) with $d_0=0$ and $d_1 \neq 0$ for an arbitrary measured value $x$, the real displacement $\alpha$ must satisfy
\begin{equation}
    H_N(\tilde{x}) = 0, \qquad
    N t_a \sqrt{2} H_{N-1} (\tilde{x}) \neq  \alpha r_a
    H_N(\tilde{x}),
\end{equation}
where $\tilde{x}=x-\alpha t_a/(\sqrt{2}r_a)$.
The first condition can be for all values of $x$ fulfilled by a suitable choice of $\alpha$, while the second condition is in these cases satisfied automatically, as Hermite polynomials of unequal orders have different roots.
To summarize, the universal setup for deterministic generation of a state (\ref{1step02}) from a completely arbitrary state $|\psi_N\rangle$ consists of a beam splitter, homodyne detection, and a suitable displacement operation, where the specific values of parameters have to be adjusted according to the state employed.
Also note that the displacement operation could be replaced by a suitable post-selection - allowing only states for which no displacement is necessary and discarding the rest. Thus, experimental feasibility can be gained at the cost of a reduced success rate.

To perform the approximate photon subtraction on the input state $|\chi_M\rangle$, this state in mode 1 is combined with vacuum in mode 2 on a beam splitter with transmittance $t_a$
yielding a two-mode entangled state at the output. A balanced beam splitter and a pair of homodyne detectors are then used to project the mode $2$ and the mode $4$ prepared in
auxiliary state $|\phi_{\overline{0}}\rangle$ onto the EPR state $|\Psi_{\mathrm{EPR}}\rangle$, c.f Fig.~1(b).
This conditionally prepares the remaining output mode $1$ in the state $|\chi_{M-1}\rangle=\sum_{m=0}^{M-1} b_m' |m\rangle$,  where
\begin{equation}
    b_m' = \sum_{k=m+1}^M d_{k-m} b_k  \sqrt{ {k \choose m}}
    t_a^m r_a^{k-m}.
\end{equation}

\begin{figure}[!t!]
\centerline{\psfig{figure=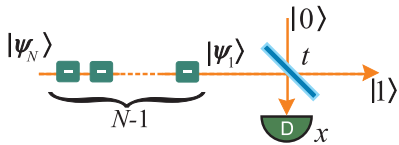,width=0.6\linewidth}}
\caption{(color online) Complete setup for generation of a single-photon state.} \label{fig_total}
\end{figure}

\section{Preparation of single-photon state}

The complete scheme for preparation of single-photon state from $N$ copies of state $|\psi_N\rangle$
is shown in Fig.~\ref{fig_total}. By repeated application of the approximate photon subtraction  we can transform any state $|\psi_N\rangle$ to a state 
\begin{equation}
    |\psi_1\rangle = a_0|0\rangle + a_1|1\rangle,
\end{equation}
with $|a_1|>0$. The parameters $a_0$ and $a_1$  can be made real by a suitable phase shift. This state is then combined with vacuum on a beam splitter with transmittance $t$, \petr{after which a homodyne detection of the amplitude quadrature $x$ of one output mode is performed, projecting the state onto}
 \begin{equation}\label{laststep}
    |\psi_{\mathrm{out}}\rangle \propto ( a_0  + \sqrt{2} x r a_1) |0\rangle + ta_1|1\rangle .
\end{equation}
If we postselect the events when the measurement outcome is $x = -a_0/(\sqrt{2}r a_1)$, we
remove the vacuum term by destructive quantum interference and  obtain the desired single-photon state.

\begin{figure}
\centerline{\psfig{figure=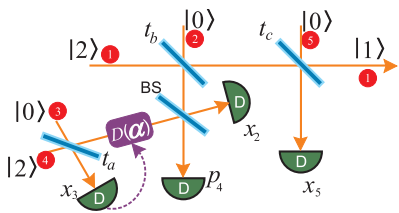,width=0.6\linewidth}}
\caption{(color online) Complete setup for generation of a single photon state from a pair of two photon states. $t_a$, $t_b$ and $t_c$ denote transmittances of the respective beam splitters, while $BS$ stands for a balanced beam splitter. Numbers 1 to 5 are used to label the modes involved.} \label{fig_twophoton}
\end{figure}

As a demonstration, let us now explicitly show the procedure to create a single-photon state from a pair of two-photon states $|2\rangle$. The full scheme is presented in Fig.~\ref{fig_twophoton}. 
It can be easily shown that to generate the single-photon state the feed-forward displacement $\alpha$ should read
\begin{equation}\label{alpha}
    \alpha = \frac{r_a}{t_a}(x_3 \sqrt{2}-1),
\end{equation}
where $x_3$ represents a value obtained by the homodyne measurement of the amplitude quadrature $x_3$ of mode 3. The other three homodyne detectors measure amplitude quadratures $x_2$ and $x_5$ of modes $2$ and $5$, respectively, and phase quadrature $p_4$ of mode $4$. Successful preparation of state $|1\rangle$ is heralded by the measurement outcomes
\begin{equation}
x_2 = 0, \quad p_4 = 0, \quad
 x_5 = - r_b \frac{t_a^2 + 2(x_3\sqrt{2} -1)r_a^2}{t_a r_a t_b r_c 2\sqrt{2}}.
 \label{x5formula}
\end{equation}

\begin{figure}[!t!]
\centerline{\psfig{figure=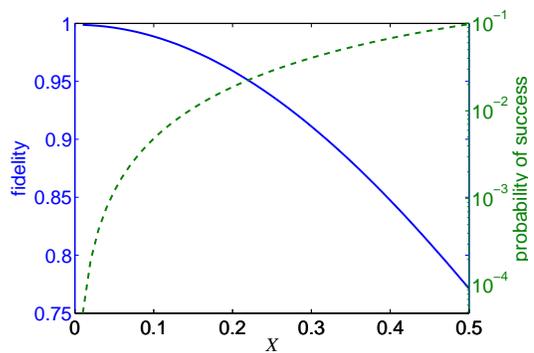,width=0.8\linewidth}}
\caption{(color online) Fidelity (left, blue solid line) and probability of success (right, green dashed line) of the preparation of the single photon state from a pair of two photon states with respect to the post-selection threshold $X$.  } \label{fig_twophotonreal}
\end{figure}

In real experimental practice we cannot condition on the projection on a single quadrature eigenstate $|x = \xi\rangle$, as this corresponds to an event with zero probability of success. Instead, we have to accept all events when the measured value falls within a narrow interval centered at $\xi$, thus realizing a POVM element
\begin{equation}\label{homodyne}
    \Pi_{k,x=\xi} = \int_{-X}^{X}|x = \xi+q\rangle_k\langle x = \xi + q| dq,
\end{equation}
where the parameter $X$ determines the half-width of the post-selection interval and $k$ labels the mode that is measured. This of course effects the output state. The global input state encompassing five modes, as can be seen in Fig.~\ref{fig_twophoton}, can be expressed as
\begin{equation}
    |\psi_{\mathrm{in}}\rangle = |2\rangle_1|0\rangle_2|0\rangle_3|2\rangle_4|0\rangle_5.
\end{equation}
After after interactions on all beam splitters and the feed-forward loop the output state for a single particular measured value $x_3$ reads
\begin{equation}
    |\psi_{\mathrm{out}}(x_3)\rangle = U_{c,15}U_{\mathrm{BS},24}U_{b,12}D_4(\alpha)\langle x_3|_3U_{a,34}|\psi_{\mathrm{in}}\rangle.
\end{equation}
Here, $U_{j,kl}$ represents a unitary transformation of a beam splitter $j = a,b,c,\mathrm{BS}$ coupling a pair of modes $k,l$, $D_4(\alpha)$ represents the displacement  performed on mode 4 and $\alpha$ is given by Eq. (\ref{alpha}).  The final state is given by
\begin{eqnarray}\label{rhofin}
    \rho_{\mathrm{1}}(x_3) &=& \frac{\Tr_{2345}[ \Pi_{2,x=0} \Pi_{4,p=0}\Pi_{5,x=x_5}|\psi_{\mathrm{out}}\rangle\langle\psi_{\mathrm{out}}|]}{P_S(x_3)},\nonumber
\end{eqnarray}
where $x_5$ is given by Eq. (\ref{x5formula}) and we have avoided to explicitly mark the dependence of $|\psi_{\mathrm{out}}\rangle$ on the value $x_3$ for the sake of brevity. $\Tr_{2345}$ stands for the partial trace over all modes other than mode 1 and $P_S(x_3)$ denotes the probability of success
\begin{equation}
    P_S(x_3) = \Tr[ \Pi_{2,x=0}\Pi_{4,p=0}\Pi_{5,x=x_5}|\psi_{\mathrm{out}}\rangle\langle\psi_{\mathrm{out}}|].
\end{equation}
This, however, still corresponds only to the scenario when a particular value $x_3$ was detected. To obtain the final result, we need to average
the state (\ref{rhofin}) over all possible experimental outcomes, arriving at
\begin{equation}\label{rhofin2}
    \rho_{1} = \frac{1}{P_S}\int_{-\infty}^{\infty} P_S(x_3) \rho_{1}(x_3) d x_3,
\end{equation}
with a probability of success
$   P_S = \int_{-\infty}^{\infty} P_S(x_3) d x_3.$

Figure~\ref{fig_twophotonreal} shows the performance of the procedure with respect to homodyne detection with nonzero threshold $X$. As the measure of quality we employ the fidelity, $F = \langle 1|\rho_{1}|1\rangle$, which in this case reliably quantifies the content of the single-photon state in the overall mixture. The transmittances of the beam splitters were optimized as to maximize the probability of success $P_S$ in the limit of very narrow acceptance windows ($X\rightarrow 0$): $t_a = 0.62$, $t_b = 0.79$ and $t_c = 0.90$. The trade-off between fidelity and the success probability 
is clearly visible in Fig.~\ref{fig_twophotonreal}.

\begin{figure}[!t!]
\centerline{\psfig{figure=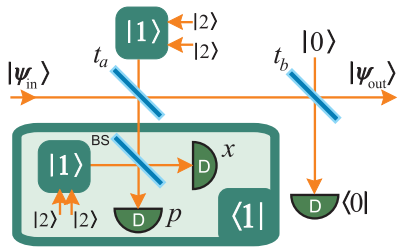,width=0.60\linewidth}}
\caption{(color online) Complete setup for implementation of the nonlinear sign gate using only Gaussian operations and two photon states as a resource. $t_a$ and $t_b$ transmittances of the respective beam splitters, while $BS$ stands for a balanced beam splitter. } \label{fig_kerrtwophotonscheme}
\end{figure}

\section{An example: nonlinear SIGN gate}

Finally we are going to present a full implementation of a non-Gaussian operation using only Gaussian operations and measurements and ancillary states $|\psi_N\rangle$. The resource states are again going to be the two-photon states $|2\rangle$ from which the single-photon states are extracted by means of procedure depicted in Fig.~\ref{fig_twophoton}. The non-Gaussian operation under consideration is the nonlinear sign gate \cite{Knill01}
which transforms a generic state $|\psi_{\mathrm{in}}\rangle = c_0 |0\rangle + c_1|1\rangle + c_2|2\rangle$ into $|\psi_{\mathrm{out}}\rangle = c_0|0\rangle + c_1|1\rangle - c_2|2\rangle$.  This represents a unitary evolution induced by a Kerr-type Hamiltonian $\hat{H}=\frac{\pi}{2}\hat{n}(\hat{n}-1)$ on a three-dimensional Hilbert space spanned by $|0\rangle,|1\rangle,|2\rangle$.

A celebrated result in linear-optics quantum computing is that this gate
can be implemented with help of only beam splitters, one ancillary single-photon state, and two measurements, one projecting on a single-photon state, the other on the vacuum state \cite{NSG}, see Fig.~\ref{fig_kerrtwophotonscheme}. The single-photon state projection can be performed with help of a scheme in Fig.~1(a) while the projection on the vacuum state is a Gaussian operation. The transmittances of the beam splitter must satisfy $t_a^2 = (3-\sqrt{2})/7 \approx 0.23$ and $t_b = t_a/(1-2t_a^2) \approx 0.87$ \cite{NSG}.

The performance of the gate can be evaluated by using the quantum process fidelity. Consider a maximally entangled state on the Hilbert space $\mathcal{H}_{N=2}^{\otimes 2}$, $|\Phi_{012}\rangle = (|00\rangle + |11\rangle + |22\rangle)/\sqrt{3}$. Applying the nonlinear sign gate on one of the modes transforms the state into
$    |\Phi'_{012}\rangle = (|00\rangle + |11\rangle - |22\rangle)/\sqrt{3}.$
With the help of this state the gate could be applied by means of teleportation to an arbitrary unknown state \cite{Gottesman99}. In this sense, the measure of quality of the state $|\Phi'_{012}\rangle$ can serve as a tool to evaluate the quality of operation.

\begin{figure}[!t!]
\centerline{\psfig{figure=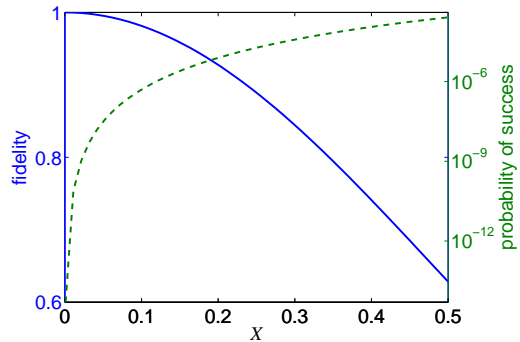,width=0.8\linewidth}}
\caption{(color online) Fidelity (left, blue solid line) and probability of success (right, green dashed line) for implementation of the nonlinear sign gate with respect to the post-selection threshold $X$.  } \label{fig_kerrtwophoton}
\end{figure}

Using similar calculations as before, we can determine the mixed two-mode state $\rho_{012}$ produced by the scheme and the success probability of the scheme for finite acceptance windows on homodyne detections. The fidelity of the operation can now be expressed as
$F = \langle \Phi'_{012}|\rho_{012}|\Phi'_{012}\rangle.$
Figure~\ref{fig_kerrtwophoton} shows the resulting relations between the fidelity, the post-selection threshold $X$, and the probability of success.

\section{Conclusions}

In summary, we have demonstrated  that a steady supply of pure non-Gaussian states possessing finite expansion in the Fock-state basis, together with the experimentally readily accessible Gaussian operations and Gaussian measurements, is sufficient for universal quantum state manipulation and engineering.  \petr{The required ancilla non-Gaussian states could be generated e.g. using squeezing operations, coherent displacements and conditional single-photon subtraction \cite{Fiurasek05}. The conditional photon subtraction can be performed reliably with avalanche photo-diode detectors even though their overall detection efficiency is of the order of 50\% or even lower. The low efficincy only reduces the success probability of the state-preparation scheme but not the fidelity of the prepared state \cite{Fiurasek05}. In contrast, such detectors are unsuitable for direct implementation of measurement induced non-Gaussian operations using the schemes proposed in Refs. \cite{CKW,Fiurasek03} becuase the low efficiency would imply reduced fidelity of the gate. In our approach we thus replace direct single-photon detection by an indirect detection relying on off-line produced non-Gaussian states and homodyne detection. In this way it is possible to achieve high fidelity at the expense of probabilistic nature of the scheme.}
Our generic scheme involves several optimization possibilities and its efficiency can be improved by tuning the  transmittances of beam splitters and the widths of the acceptance windows of homodyne measurements. Moreover, it is likely that for each particular task the efficiency can be improved significantly by using  specific dedicated scheme tailored to a given resource state $|\psi_N\rangle.$

 \petr{Besides states with finite Fock-state expansion also other classes of states could be sufficient for universal CV quantum gate engineering. However, dealing with completely generic states in infinite-dimensional Hilbert space of the quantized electromagnetic field is extremly difficult due to the \emph{a-priori} infinite number of parameters. It is unlikely that the question of sufficiency of a given state for universal CV quantum gate engineering could be decided in a completely general way. Instead, partial ad-hoc solutions could be provided for certain finite-parametric classes of states (e.g. the cubic phase state proposed in Ref. \cite{Gottesman01}). Identifying such potentially useful classes of states is an interesting open problem which, however, is beyond the scope of our present work.}

Our findings shall find applications in advanced optical quantum information processing and quantum state engineering. On more fundamental side, our results shed more light on the quantum information processing power of non-Gaussian states  and they help to bridge the gap between single-photon and continuous-variable approaches.

\begin{acknowledgments}
This work was supported by MSMT under projects LC06007, MSM6198959213, and 7E08028,
and also by the EU under the FET-Open project COMPAS (212008).
\end{acknowledgments}

\end{document}